\documentclass[aip,amsmath,amssymb,reprint]{revtex4-2}

\usepackage[final]{graphicx}
\usepackage{float}
\usepackage{hyperref}
\usepackage{amsmath}
\usepackage[normalem]{ulem}
\hypersetup{colorlinks=true, citecolor=blue, urlcolor=blue, linkcolor=blue}
\usepackage{color}

\newcommand{\BSCCO}{Bi$_2$Sr$_2$CaCu$_2$O$_{8+\delta}$}

\makeatletter
\def\@email#1#2{%
 \endgroup
 \patchcmd{\titleblock@produce}
  {\frontmatter@RRAPformat}
  {\frontmatter@RRAPformat{\produce@RRAP{*#1\href{mailto:#2}{#2}}}\frontmatter@RRAPformat}
  {}{}
}%
\makeatother

\begin{document}

\author{Nicolas Gauthier}
\email{nicolas.gauthier@inrs.ca}
\affiliation{Advanced Laser Light Source, Institut National de la Recherche Scientifique - Énergie Matériaux Télécommunications Varennes QC J3X 1P7 Canada}
\author{Benson Kwaku Frimpong}
\affiliation{Advanced Laser Light Source, Institut National de la Recherche Scientifique - Énergie Matériaux Télécommunications Varennes QC J3X 1P7 Canada}
\author{Dario Armanno}
\affiliation{Advanced Laser Light Source, Institut National de la Recherche Scientifique - Énergie Matériaux Télécommunications Varennes QC J3X 1P7 Canada}
\affiliation{Department of Physics, Center for the Physics of Materials, McGill University, 3600 rue Université, Montréal, Québec H3A 2T8, Canada}
\affiliation{Department of Chemistry, McGill University, 801 rue Sherbrooke Ouest, Montréal, Québec H3A 0B8, Canada}
\author{Akib Jabed}
\affiliation{Advanced Laser Light Source, Institut National de la Recherche Scientifique - Énergie Matériaux Télécommunications Varennes QC J3X 1P7 Canada}
\author{Francesco Goto}
\affiliation{Advanced Laser Light Source, Institut National de la Recherche Scientifique - Énergie Matériaux Télécommunications Varennes QC J3X 1P7 Canada}
\author{Vicky Hasse}
\affiliation{Max Planck Institute for Chemical Physics of Solids, 01187 Dresden, Germany}
\author{Claudia Felser}
\affiliation{Max Planck Institute for Chemical Physics of Solids, 01187 Dresden, Germany}
\author{Genda Gu}
\affiliation{Condensed Matter Physics and Materials Science,
Brookhaven National Laboratory, Upton, NY 11973, USA}
\author{Heide Ibrahim}
\affiliation{Advanced Laser Light Source, Institut National de la Recherche Scientifique - Énergie Matériaux Télécommunications Varennes QC J3X 1P7 Canada}
\author{Francois Légaré}
\affiliation{Advanced Laser Light Source, Institut National de la Recherche Scientifique - Énergie Matériaux Télécommunications Varennes QC J3X 1P7 Canada}
\author{Fabio Boschini}
\email{fabio.boschini@inrs.ca}
\affiliation{Advanced Laser Light Source, Institut National de la Recherche Scientifique - Énergie Matériaux Télécommunications Varennes QC J3X 1P7 Canada}
\affiliation{Quantum Matter Institute, University of British Columbia, Vancouver, BC V6T 1Z4, Canada}

\title{Detecting the full photoemission cone from laser-based ARPES experiments by leveraging deflector technology}

\date{\today}

\begin{abstract}
Angle-resolved photoemission spectroscopy (ARPES) provides a direct access to the electronic band structure of solid and molecular systems. The momentum range accessible by this technique depends directly on the photon energy used, and low-photon-energy sources are insufficient to photoemit electrons over the full Brillouin zone of most quantum materials. 
In addition, while electrons are emitted over a $2\pi$ solid angle, conventional hemispherical analyzers only collect a small subset of those electrons. A previous work [RSI 92, 123907 (2021)] demonstrated that electrons emitted over a larger field-of-view can be acquired in one fixed configuration by accelerating them towards the analyzer with a bias voltage. 
Here, we extend this work by leveraging the deflector technology of novel ARPES hemispherical analyzers. We demonstrate the ability to detect all $2\pi$ photoemitted electrons in a fixed configuration for various materials such as gold, cuprates and transition-metal dichalcogenides. This approach is especially advantageous for time-resolved ARPES, as electron dynamics over a large momentum range can be accessed with identical measurement conditions.
\end{abstract}
\maketitle

\section{Introduction}

Angle-resolved photoemission spectroscopy (ARPES) is a well-established technique in condensed matter physics that grants direct access to the electronic band structure and many-body interactions in quantum materials.\cite{Sobota2021,damascelli2004probing}  Photoemission is a photon-in-electron-out process, and the choice of light source and photon energy determines many working parameters, such as the surface-to-bulk sensitivity, the energy resolution, and the momentum space coverage. Furthermore, the advent of ultrafast laser light sources enabled the development of time-resolved ARPES (TR-ARPES), which is now a well-established momentum-resolved pump-probe technique to investigate and control quantum materials.\cite{Boschini2024} 

\begin{figure*}
\includegraphics[scale=1]{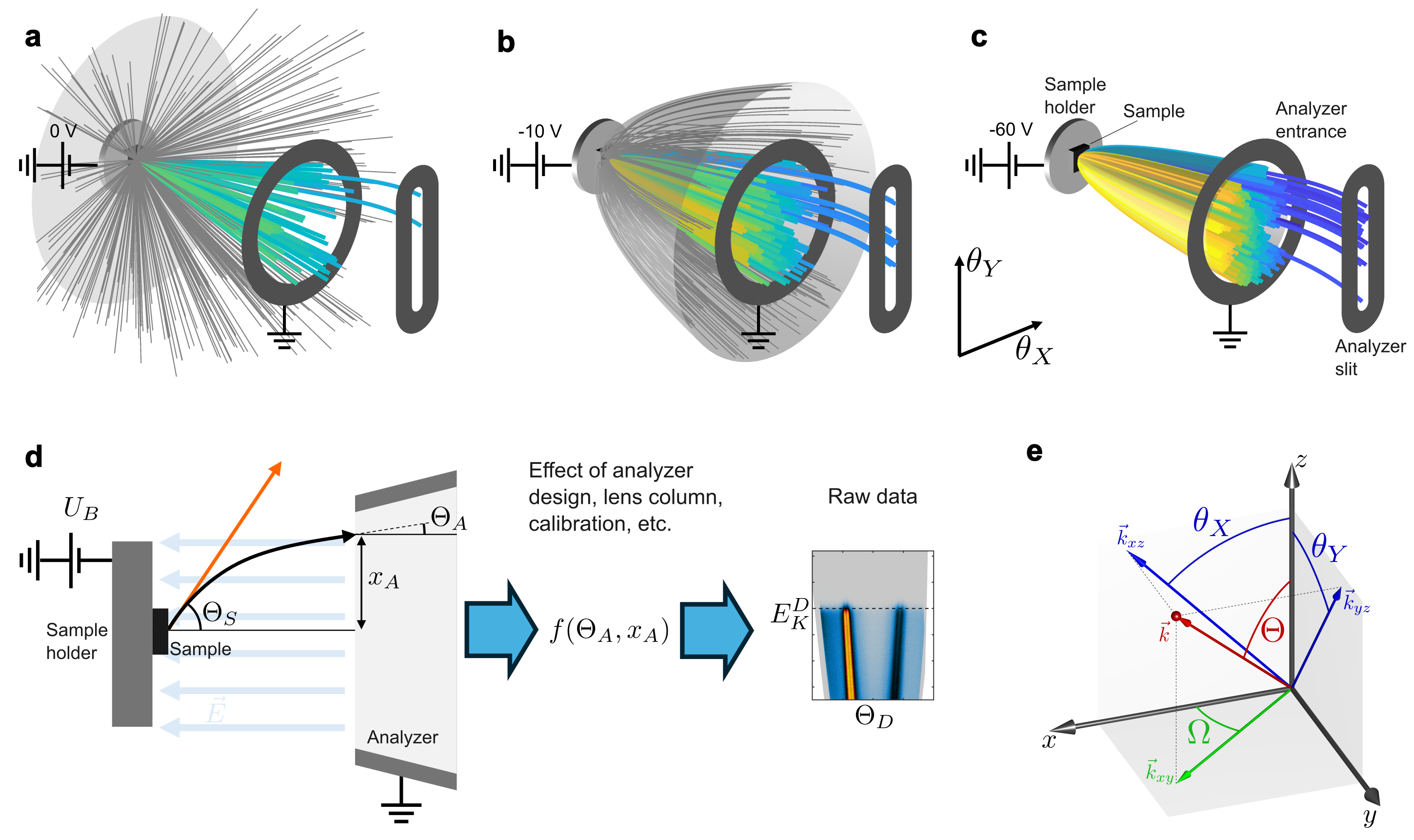}
\caption{
(a-c) Illustration of electron trajectories with the application of different effective bias voltages ($U_B=0$, -10 and -60~V). Electrons are emitted over a $2\pi$ solid angle (delimited by the gray plane), but only the electrons entering the analyzer are colored, with a color gradient reflecting their $k_x$ value. The analyzer deflector deviates those electrons and, in combination with the slit, allows to select which $\theta_X$ (associated to $k_x$) is measured. Here, the positive-$\theta_X$ electrons (blue) are accepted through the slit after being deflected (in a schematically simplified way). In the scenario at $U_B=-60$~V, all photoemitted electrons enter the analyzer and can be measured by scanning the deflector angle.  
(d) An electron emitted from the sample at an angle $\Theta_S$ propagates in a straight line in the absence of effective bias voltage, as illustrated by the orange arrow. With a voltage bias $U_B$ applied to the sample, the electron is accelerated by the electric field $\vec{E}$ towards the analyzer entrance, arriving at a position $x_A$ and an angle $\Theta_A$. The exact trajectory inside the analyzer depends on these variables and the electron is finally measured at an angle $\Theta_D$ on the detector. 
(e) Projections of the electron momentum $\vec{k}$ in the $xy$, $xz$ and $yz$ planes to define the measured photoemission angles $\theta_X$ and $\theta_Y$ and the angles converted in spherical coordinates $\Theta$ and $\Omega$.
}
\label{fig:Schematic}
\end{figure*}

While synchrotron facilities can provide widely tunable photon energies for conventional ARPES measurements, the probe light for TR-ARPES is generally constrained by the range of frequencies that can be produced with ultrafast lasers.\cite{Na2023, gauthier2020tuning}
Low-photon energy ($\sim 6$~eV) TR-ARPES systems are commonly used because of their ease of implementation and reliability. However, a significant drawback for 6~eV probe is its limited momentum coverage. Indeed, the accessible momentum range is directly related to the probe photon energy and the Brillouin zone boundary of most materials is out of reach. 
The inherent momentum limitation of low-photon energy ARPES systems is further aggravated when using conventional hemispherical analyzers, since they only accept a small subset of the electrons emitted over the complete $2\pi$ solid angle. Indeed, the entrance aperture of most electron analyzers have an acceptance angle of only $\pm15^\circ$ relative to the analyzer axis, corresponding to $\sim 25\%$ of the complete solid angle emitted. In addition, the entrance slit in front of the hemispherical capacitor further reduces the photoelectron acceptance to only a narrow slice (see Fig.\,\ref{fig:Schematic}a). 

The angular acceptance limitation can be partially overcome through the acceleration of electrons towards the analyzer with an electric field, as shown by Gauthier \textit{et al.}\cite{Gauthier2021} With enough acceleration, all the electrons emitted over the $2\pi$ solid angle enter the analyzer nose and a complete slice ($\pm90^\circ$) is acquired in one fixed configuration. 
While this bias-voltage approach does not overcome the physical emission limit imposed by the low photon energy, it greatly expands the flexibility of laser-based ARPES setups. This experimental strategy has been implemented in many ARPES systems\cite{Yamane2019,Gauthier2021,OLeary2024,Yang2024,Majchrzak2024}, and applied to investigate transition-metal dichalcogenides,\cite{Gauthier2025}
topological insulators,\cite{Gauthier2024,jabed2025control}
charge density wave systems,\cite{OLeary2024,Gauthier2021}
nickelates~\cite{Yang2024}
and cuprates.\cite{Armanno2025}

Unfortunately, the approach described in Ref.\,\onlinecite{Gauthier2021} is only applicable when the sample surface is normal to the hemispherical analyzer axis, and the complete photoelectron $2\pi$ solid angle can only be obtained by mechanically changing the sample azimuthal angle. Advantageously, state-of-the-art hemispherical analyzers equipped with deflector technology remove the need for sample rotation by measuring the photoelectron emitted not only along the slit orientation but also perpendicular to it.~\cite{Ishida2018} In this work, we combine the bias voltage approach with the deflector technology and demonstrate the acquisition of all $2\pi$-photoemitted electrons in a fixed configuration (see Fig.\,\ref{fig:Schematic}c). In Section~\ref{sec:model}, we first extend the previously established model to include the deflector axis and present the experimental details in Section~\ref{sec:expdetails}. In Section~\ref{Sec:Results}, the technical validity of this extension is demonstrated and characterized with measurements performed on Au(111), and further showcased with results on the high-temperature superconductor \BSCCO\ and the transition-metal dichalcogenide WTe$_2$.

\section{Extension of the model for hemispherical analyzers with deflectors}
\label{sec:model}
\subsection{Established model for hemispherical analyzers}
\label{subsecOldModel}

First, we recall the considerations, assumptions and equations of the previously established model~\cite{Gauthier2021} and include a schematic illustration with the important variables in Fig.\,\ref{fig:Schematic}d. We consider photoelectrons emitted from a sample whose normal is parallel to the analyzer axis (i.e. normal emission geometry). The bias voltage is applied to the sample, and the analyzer nose is grounded. The measured kinetic energy at the detector $E_K^D$ is increased relative to the one at the sample surface $E_K^S$ by the effective bias voltage $U_B^*$
\begin{equation}
    E_K^S=E_K^D+eU_B^*,
\end{equation}
where the effective bias voltage is defined as the applied bias voltage $U_B$ and the workfunction difference between the sample and the analyzer
\begin{equation}
    U_B^*=U_B + (\Phi_A-\Phi_S)/e.
\end{equation}

We assume a uniform electric field between the sample and the analyzer nose (parallel plate capacitor approximation), which is a reasonable assumption in the normal emission geometry. This allows us to analytically evaluate the photoelectron trajectories from emission to their entrance into the analyzer nose. Inside the analyzer, the electron trajectories are significantly more complex. Nevertheless, it can be generally assumed that the detected angle $\Theta_D$ is a function of the photoelectron position $x_A$ and angle $\Theta_A$ at the analyzer entrance
\begin{equation}
\Theta_D = f (\Theta_A,x_A),
\label{EQ_AngleDet}
\end{equation}
as sketched in Fig.\,\ref{fig:Schematic}d.
Since the analytical form of Eq.\,\ref{EQ_AngleDet} is a priori unknown, we focus on the two limit cases that can be treated analytically: the angular limit $\Theta_D \approx f (\Theta_A)$ and the position limit $\Theta_D \approx f (x_A)$. As described in details previously~\cite{Gauthier2021} {and summarized in the appendix~\ref{appendix:kconversion},} the angular-to-momentum conversion is given by
\begin{equation}
k_\parallel^S=\frac{1}{\hbar} \sqrt{2mE_K^S} F \sin \Theta_D,
\label{EQ_angle2mom}
\end{equation}
where $F$ is a scaling factor that depends on the chosen limit. For the angular limit, we have
\begin{equation}
F_A=\sqrt{1+2\alpha} ,
\end{equation}
while the position limit gives
\begin{equation}
F_P= \sqrt{\frac{\alpha+1+\sqrt{2 \alpha + 1 - \alpha^2 \tan^2 \Theta_D}}{2}},
\end{equation}
where the kinetic parameter $\alpha$ is
\begin{equation}
\alpha=\frac{-eU_B^*}{2E_K^S}.
\label{eqAlpha}
\end{equation}
In addition, the low-energy cutoff (LEC), i.e. the cutoff below which photoelectrons cannot exit the sample due to geometry or energetic constraints, can be written for both limits as 
\begin{equation}
E_{K,LEC}^D = -e U_B^* \left[ 1 + \left(
\frac{\tan \Theta_D^{LEC}}{\gamma}
\right)^2 \right],
\label{EQ_LEC}
\end{equation}
with $\gamma=1$ and 2 for the angular and position limits, respectively. Comparing the measured LEC with the model expectation is a simple approach to evaluate which limit is more appropriate. 

\subsection{Extended model for analyzers with deflectors}

In conventional hemispherical analyzers, only electrons entering the slit are detected, and we label their emission angle along the slit direction $\theta_Y$. The emission angle along the transverse direction, labeled $\theta_X$, is experimentally constrained to be zero. The deflectors of novel analyzers allow for sweeping through a range of $\theta_X$, providing access to another momentum axis without mechanically moving the sample. 
It is convenient to rewrite the measured photoemission angles $\theta_X$ and $\theta_Y$ as polar and azimuthal angles $\Theta$ and $\Omega$, respectively (see Fig.\,\ref{fig:Schematic}e). In such notation, the conventional angle-to-momentum conversion equation reads
\begin{equation}
k_\parallel = \sqrt{2mE_K}\cos \Theta,
\label{eq_comm1}
\end{equation}
\begin{equation}
(k_x,k_y) =  (k_\parallel \cos \Omega, k_\parallel \sin \Omega).
\label{eq_comm2}
\end{equation}
Note that these two angular notations ($\theta_X$,$\theta_Y$)$\leftrightarrow$($\Theta$,$\Omega$) are commonly related using  $\Theta^2=\theta_X^2+\theta_Y^2$ and $\tan \Omega = {\theta_Y}/{\theta_X}$.~\cite{Ishida2018} However, these equations are not strictly correct since these trigonometric relations apply to spatial dimensions and not angles. The formally correct approach is to define a vector representing the photoelectron emission direction, and evaluate its projections in the $xz$ and $yz$ planes, with $\theta_X$ and $\theta_Y$ being the angles between these projections and the $z$-axis. This approach leads to the following relations:
\begin{equation}
    \tan^2 \Theta = \tan^2 \theta_X + \tan^2 \theta_Y
    \label{eq:thetaAngle}
\end{equation}
and
\begin{equation}
    \tan \Omega = \frac{\tan \theta_Y}{\tan \theta_X}.
    \label{eq:omegaAngle}
\end{equation}
The commonly used equations ($\Theta^2=\theta_X^2+\theta_Y^2$ and $\tan \Omega = {\theta_Y}/{\theta_X}$) are a good approximation of these exact equations for $\theta_X$ and $\theta_Y$ below $\sim20^\circ$. Beyond this range, exact equations Eqs.\,\ref{eq:thetaAngle} and \ref{eq:omegaAngle} should always be used for accuracy. As most hemispherical analyzers have an angular acceptance of $\pm15^\circ$, both sets of equations are generally valid. This is also true in the presence of a bias voltage on the sample, since these angles refer to the detected angles and not the emission angle at the surface. However, note that Eqs.\,\ref{eq:thetaAngle} and \ref{eq:omegaAngle} are used throughout all this work for maximal accuracy.

We reiterate that all the equations in section~\ref{subsecOldModel} were originally derived assuming that $\Theta_D$ is measured along the slit direction ($\Theta_D=\theta_Y$, $\theta_X=0$). When employing hemispherical analyzers with deflectors, the component transverse to the slit ($\theta_X$) must also be considered. Cylindrical symmetry can be assumed between the sample and the analyzer nose, thereby maintaining the validity of the equations describing the photoelectron trajectories until the analyzer nose. However, the equivalence along the slit and deflector axes is not known a priori when considering the electron propagation within the analyzer, but our experimental results in section~\ref{Sec:Results} demonstrate it. 
As a consequence, the previously established conversion equation with voltage bias (Eq.\,\ref{EQ_angle2mom}) remains valid to convert the polar angle $\Theta_D$ measured using an analyzer with deflectors into the momentum $k_\parallel^S$. The azimuthal angle $\Omega_D$ only defines its $x$ and $y$ components according to Eq.\,\ref{eq_comm2}.

Furthermore, while the previous work\cite{Gauthier2021} concluded that the position limit [$\Theta_D \approx f (x_A)$] worked in all conditions for the R4000 hemispherical analyzer by Scienta Omicron, here we consider a more general possibility by taking a weighted average of the scaling factors 
\begin{equation}
F=(1-\beta)F_A + \beta F_P,
\label{eq:weight}
\end{equation}
where $\beta$ is the weighting factor. This weighted average approach is justified as $F_A$ and $F_P$ are monotonic as a function of $\alpha$, with a qualitatively similar dependence (see Fig. 2e from Ref.~\onlinecite{Gauthier2021}). This simple approach allows us to maintain the momentum conversion fully analytical for intermediate cases where both limits fail.

Finally, we also recall that the angle-to-momentum conversion leads to a renormalization of the photoemission intensity $N(E_K^S,\theta_X,\theta_Y)$ to $N(E_K^S,k_x^S,k_y^S)$ due to the change of coordinates. This relation is generalized in this work for the weighted scaling factor $F$ of Eq.\,\ref{eq:weight} and the deflector angle:
\begin{equation}
\frac{N(E_K^S,k_x^S,k_y^S)}{\Delta k_x^S \Delta k_y^S}=
\frac{N(E_K^S,\theta_X,\theta_Y)}{\Delta \theta_X \Delta \theta_Y} \left| J_{\Theta_D,\Omega_D}^{\theta_X,\theta_Y} \right|^{-1},
\end{equation}
where $\Delta k_x^S$, $\Delta k_y^S$, $\Delta \theta_X$ and $\Delta \theta_Y$ are bin sizes and $J_{\Theta_D,\Omega_D}^{\theta_X,\theta_Y}$ is the Jacobian. 
The complete intensity scaling expression is given in Appendix~\ref{appendix:IntensityScaling}.

\begin{figure*}
\centering
\includegraphics{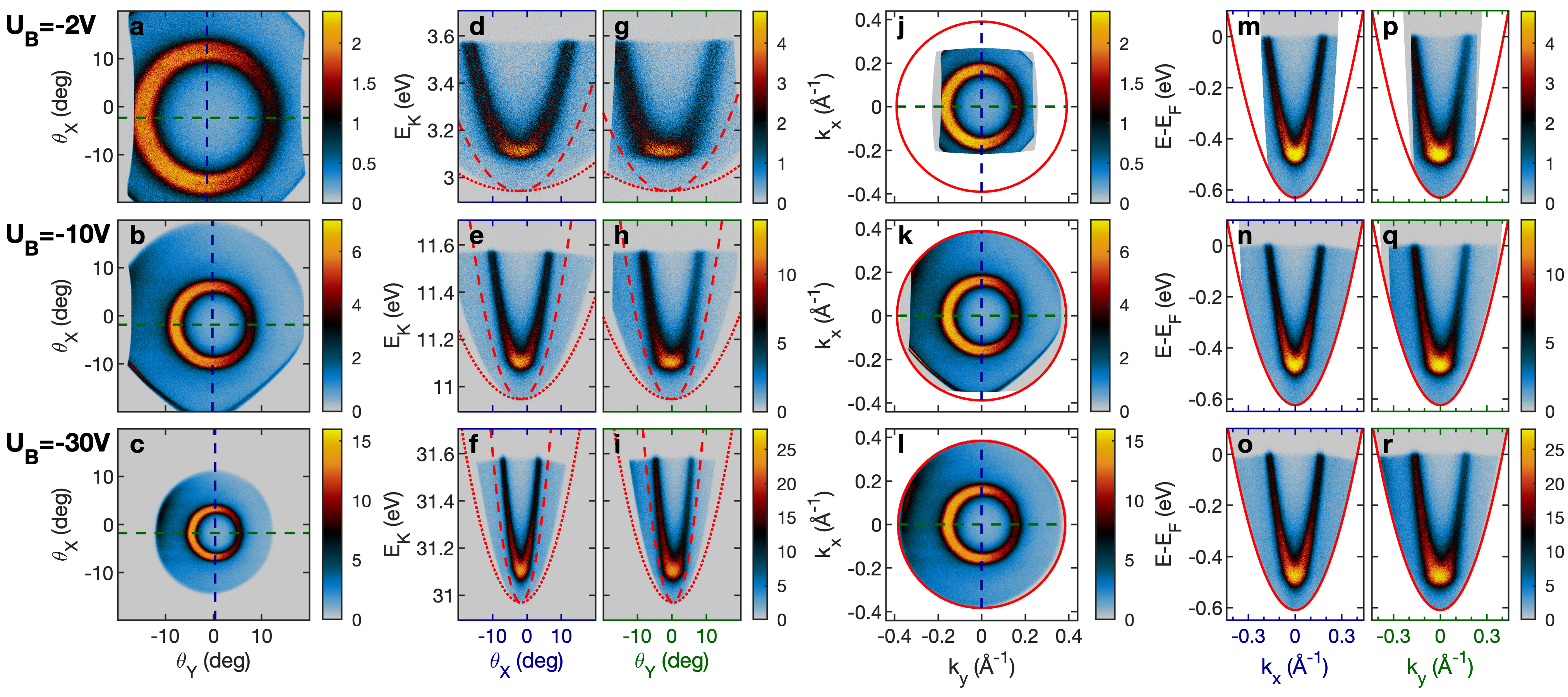}
\caption{Angle-to-momentum conversion for Au(111). (a-c) Isoenergy maps in the $\theta_X-\theta_Y$ plane at $E=E_F$ for $U_B=-2,-10$ and $-30$~V, respectively. (d-f) Spectra along $\theta_X$ for the cut marked by the dashed blue line in the corresponding panels a-c. (g-i) Spectra along $\theta_Y$ for the cut marked by the dashed green line in the corresponding panels a-c. The dashed and dotted red lines in panels (d-i) correspond to the LEC in the angular and position limits, respectively. Panels (j-r) correspond directly to panels a-i, but converted to momenta $k_x$ and $k_y$. The solid red line is the physical LEC curve. We note that regions that do not follow the LEC in panel k at negative $k_x$ are due to distortions arising from imperfect alignment. These distortions vanish at sufficiently large bias voltage. }
\label{fig:RawVSConverted}
\end{figure*}

\section{Experimental details}
\label{sec:expdetails}   

All the measurements were performed at the TR-ARPES endstation located at the Advanced Laser Light Source (ALLS) national facility. The 6 eV light source for photoemission is generated with a Yb-laser, Tangerine from Amplitude.~\cite{Longa2024} ARPES measurements were performed in ultra-high vacuum conditions, at a pressure better than $8\times10^{-11}$~mbar. This system uses the ASTRAIOS 190 hemispherical analyzer from SPECS GmbH, equipped with novel deflector technology. The analyzer slit is in the horizontal plane. The {azimuthal} sample stage is electrically decoupled from ground, and we apply bias voltage on the sample using a Hewlett-Packard 6228B dual DC power supply. 

The general characterization of the biasing method was performed on {a 1~cm diameter} Au(111) {single crystal, mounted on a flag-style sample holder}. The sample was prepared through multiple cycles of sputtering and annealing. Measurements presented in Figs.~\ref{fig:RawVSConverted}-\ref{fig:EDC} were performed at 11~K with a pass energy of 10~eV. Measurements presented in Figs.~\ref{fig:slit} and \ref{fig:flux}, on a higher quality Au(111) surface, were performed at 32~K and with pass energies of 10\,eV and 4\,eV, respectively. Samples of WTe$_2$ and \BSCCO\ were also measured to demonstrate the technique in various conditions. Those samples {were mounted on stainless steel flag-style sample plates with silver epoxy and} were cleaved in situ in ultra-high vacuum. The WTe$_2$ and \BSCCO\ samples were measured at 10~K with a pass energy of 10\,eV and 5\,eV, respectively.

The ASTRAIOS 190 analyzer has three angular modes: the medium, wide, and superwide modes with angular acceptances of $\pm10^\circ$, $\pm20^\circ$ and $\pm30^\circ$, respectively. The wide angular mode was used for all data presented in this work, as it provides a larger field-of-view than the medium angular mode at the same bias voltage and is easier to align than the superwide angular mode. Furthermore, the wide angular mode leads to a circular distribution in the $\theta_X$-$\theta_Y$ plane, while this distribution is ellipsoidal in the medium and superwide angular modes, indicating the loss of cylindrical symmetry and non-equivalence of $\theta_X$ and $\theta_Y$ in those modes.

\section{Experimental results}
\label{Sec:Results}

\subsection{General characterization}
\label{subsec:characterization}

Figure~\ref{fig:RawVSConverted} presents the measured spectra of Au(111) for voltage bias from $-2$~V to $-30$~V where surface states with circular symmetry are observed near $\Gamma$. Panels a-c present the isoenergy maps as a function of $\theta_X$ and $\theta_Y$ at the Fermi energy for different voltage values. The surface states appear at smaller angular values with increasing bias voltage. At sufficiently large bias voltage, a circular cutoff is observed beyond which no electrons are detected, and it corresponds to the LEC. Panels d-f show the ARPES spectra along $\theta_X$ with $\theta_Y=0$ for different voltages, while panels g-i present the same along $\theta_Y$ with $\theta_X=0$. 
On these panels, the LEC for the angular and position limits are shown in dashed and dotted red lines, respectively. Both LEC curves fail to capture the experimental LEC at all voltages and measurement directions, and this observation supports the use of the weighted scaling factor $F$ (Eq.~\ref{eq:weight}) for the momentum conversion. 

\begin{figure}
\centering
\includegraphics{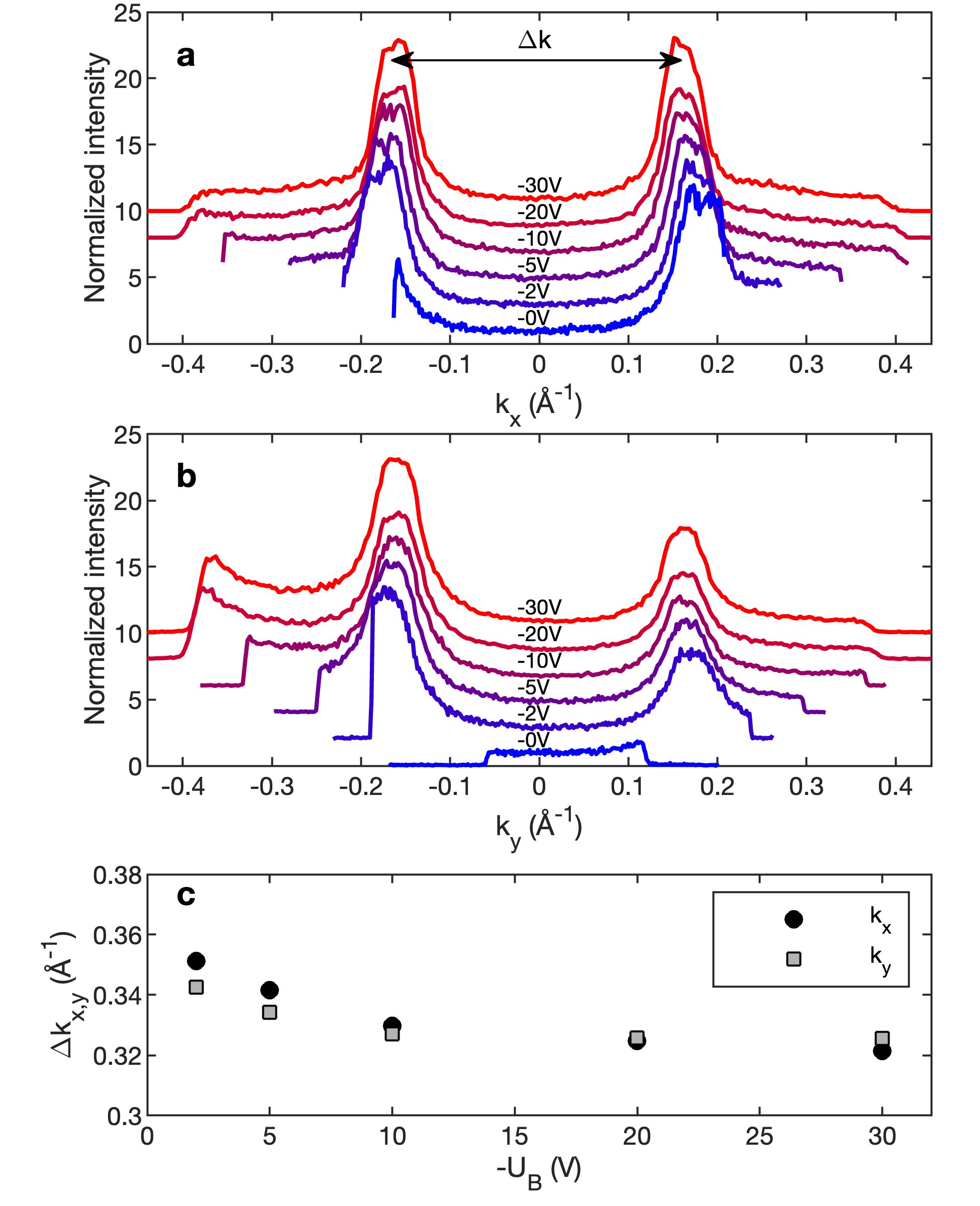}
\caption{(a) Momentum distribution curves (MDCs) of Au(111) along $k_x$ (perpendicular to the slit) taken at $E=E_F$ for $U_B$ from 0 to $-30$~V. (b) MDCs along $k_Y$ (parallel to the slit) in the same conditions. (c) Relative distance between the peaks at positive and negative momenta ($\Delta k$, as illustrated in panel a) as a function of $U_B$ for both momentum directions.}
\label{fig:MDC}
\end{figure}

\begin{figure}
\centering
\includegraphics{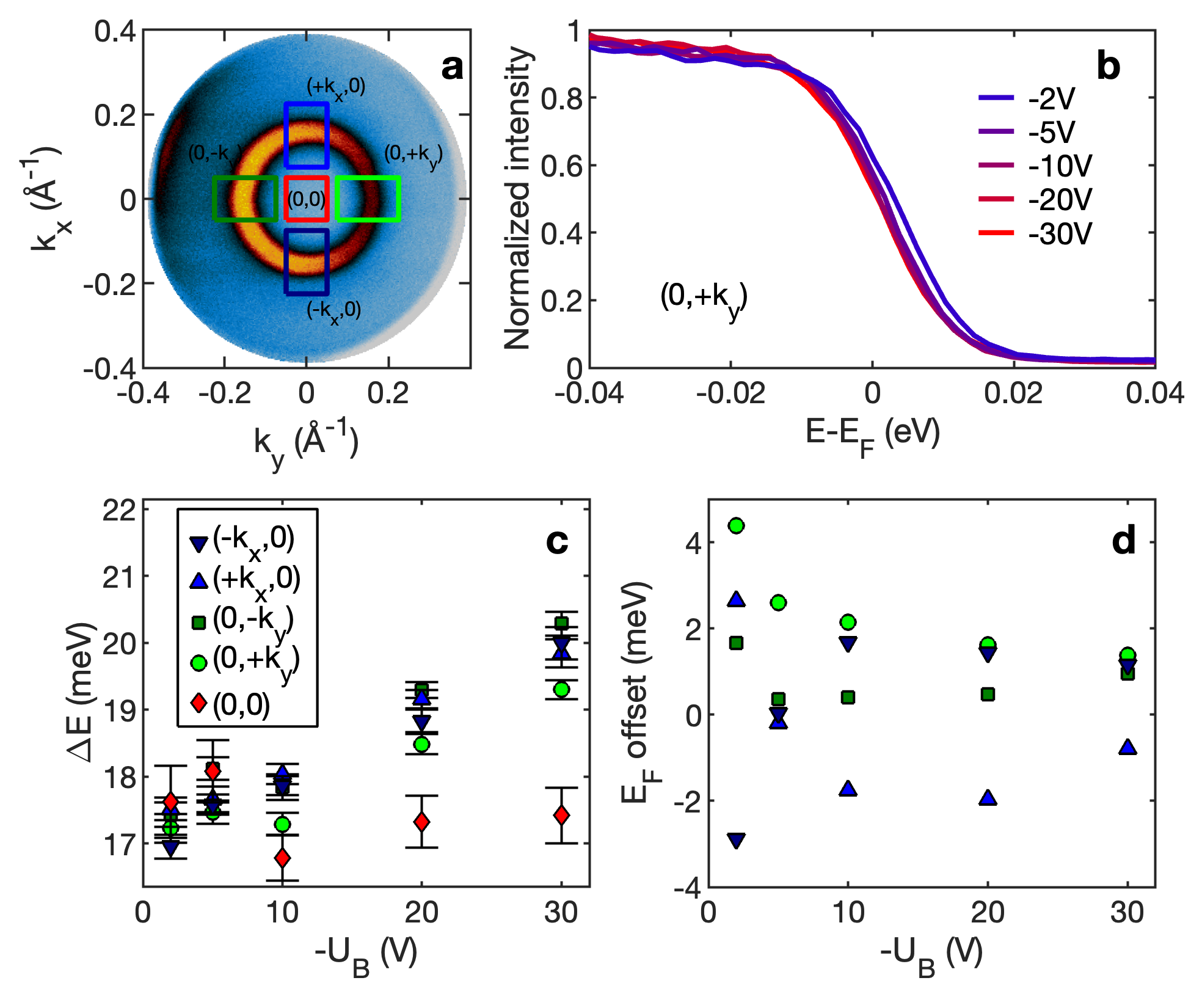}
\caption{(a) Fermi surface measured of Au(111) for $U_B=-30$~V. The {blue, green and red} boxes, labeled with their center momentum, indicate integration areas used for the EDC analysis. (b) EDCs for the $(0,+k_y)$ integration region as function of $U_B$. (c) Energy resolution determined from the EDC analysis for different momentum regions and $U_B$ values. (d) Offset of the Fermi level position determined from the EDC analysis for different momentum regions and $U_B$ values.
}
\label{fig:EDC}
\end{figure}

The full dataset in Fig.\,\ref{fig:RawVSConverted} was converted to momentum using $\beta=0.85$ and is presented in panels j-r, where the physical LEC (defined $E_K^S=\hbar^2k_\parallel^2/2m$) is indicated by the solid red line. The Fermi surfaces and ARPES spectra are voltage-independent, and their edges are in good agreement with the LEC for all momenta and energies. The converted data clearly shows that the voltage bias approach can be effectively applied to data measured with a hemispherical analyzer with deflectors. Moreover, the $\beta$ factor is voltage-independent,  providing a robust conversion without the need for fine-tuning at each bias voltage. A quantitative analysis as a function of momentum and energy is presented in Figs.~\ref{fig:MDC} and \ref{fig:EDC}. 

Momentum distribution curves (MDCs) taken along $k_x$ and $k_y$ at the Fermi level (blue and green dashed lines in Fig.\,\ref{fig:RawVSConverted}j-l) are shown in Fig.~\ref{fig:MDC}a,b for bias voltage ranging from 0 to $-30$~V. A weak deviation of the peak position as a function of voltage is observed and is quantified by extracting the relative distance $\Delta k$ between the peaks at positive and negative momenta, as shown in Fig.~\ref{fig:MDC}c. This deviation is strongest at low voltage values{, where misalignment effects are most significant}, while it saturates at large voltages. 
The peak distance $\Delta k$ along the slit ($k_y$) and transverse to it ($k_x$) are in very good agreement over the full voltage range, reinforcing the validity of our assumption to apply this model to hemispherical analyzers with deflectors.

To quantify the effect of voltage bias on the energy axis, energy distribution curves (EDCs) were taken in different momentum regions, as illustrated by the blue and green boxes in Fig.~\ref{fig:EDC}a. Representative EDCs taken in the {green} box, labeled $(0,+k_y)$, are presented in Fig.~\ref{fig:EDC}b for different $U_B$ values. Fermi edges for all $U_B$ and momentum regions are fitted by a Fermi function convoluted with a Gaussian representing the resolution function. The energy resolution extracted for the Fermi edge fits is presented in Fig.~\ref{fig:EDC}c. The energy resolution  remains the same throughout the momentum regions, but slightly increases with increasing $U_B$ due to bias-enhanced space-charge effects as discussed in the next section. {The energy resolution of the diffuse background at normal emission (red box and diamonds markers in Fig.~\ref{fig:EDC}a and \ref{fig:EDC}c, respectively) does not display any space-charge effects because of it weaker photoemission intensity.} The determined Fermi level offsets, relative to $E_F$ measured at $k_x=k_y=0$, are shown in Fig.~\ref{fig:EDC}d. A comparable spread is observed at low and high $U_B$ values, indicating that the bias is not at its origin. This spread is likely due to a misalignment of the photoemission spot with respect to the focus of the analyzer. Indeed, the ASTRAIOS 190 analyzer is very sensitive to the alignment but we observe that this limitation is effectively relaxed in the presence of applied bias voltage. 

\subsection{Additional considerations: Slit and Space Charge}
\label{subsec:furthercons}

\begin{figure}
\centering
\includegraphics{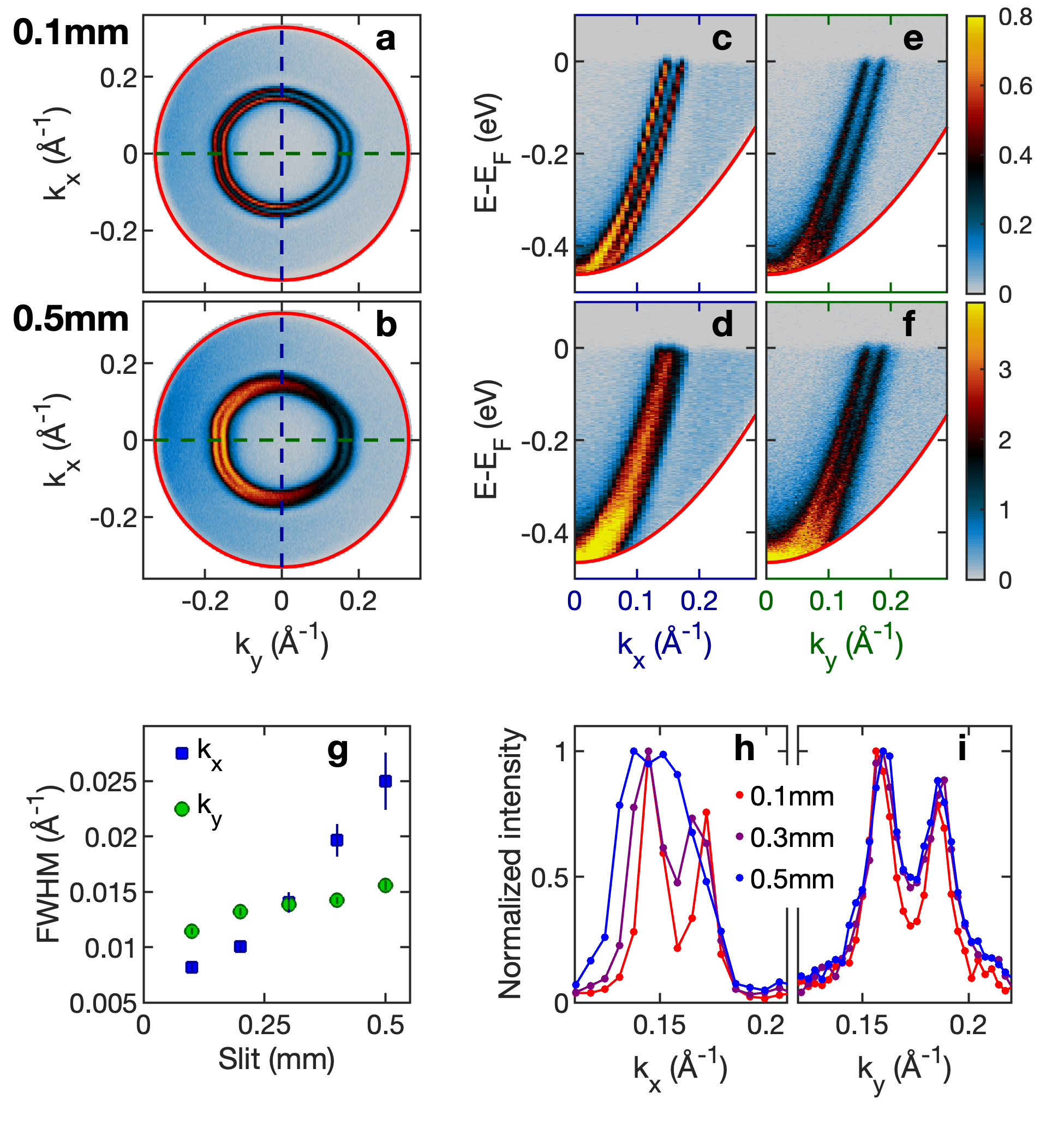}
\caption{(a,b) Fermi surface of Au(111) obtained with $U_B=-30$~V, and with slits of 0.1~mm and 0.5~mm, respectively. The spectra for each slit dimension are shown in panels (c-d) for cuts along the deflector direction ($k_x$) and in panels (e-f) for cuts along the slit direction ($k_y$). The solid red line in panels (a-f) corresponds to the physical LEC. (g) Full width at half maximum (FWHM) of the surface states at $E=E_F$ obtained along $k_x$ and $k_y$ as a function of slit dimension. (h,i) MDCs at $E=E_F$ along $k_x$ and $k_y$ for slits of 0.1, 0.3 and 0.5~mm.}
\label{fig:slit}
\end{figure}

In hemispherical analyzers, the slit size is frequently selected based on energy resolution considerations. Nevertheless, its size also affects the momentum resolution transverse to the slit (along $k_x$ in the current notation). Indeed, the larger the slit, the larger the integration range along the transverse momentum direction, thereby worsening the resolution. Applying a bias voltage has the unfortunate outcome of amplifying this resolution loss, since electrons with an even larger range of transverse momentum are accepted through the slit. Here, we illustrate this consequence 
in Fig.~\ref{fig:slit} on a pristine Au(111) where the Rashba-split surface states are well resolved.\cite{LaShell1996} Panels a and b show the Fermi surface acquired with $U_B=-30$~V, and with slits of 0.1~mm and 0.5~mm, respectively. The features along $k_x$, the direction transverse to the slit, are significantly broadened for the larger slit, while we report a minor worsening of the momentum resolution along the slit direction ($k_y$). This is also clearly seen in the spectra in panels c-f and the MDCs taken at $E_F$ presented in panels h,i. To characterize this further, MDC peaks were fitted with a Lorentzian function, and Fig.~\ref{fig:slit}g displays the full-width-at-half-maximum (FWHM) as a function of slit size.
Along the slit direction, the width increases weakly with slit size, whereas along the deflector direction, it displays a marked slit-dependence. Interestingly, we observe that the momentum resolution achieved along the deflector direction can be better than that along the slit direction for sufficiently small slits.


\begin{figure}
\centering
\includegraphics{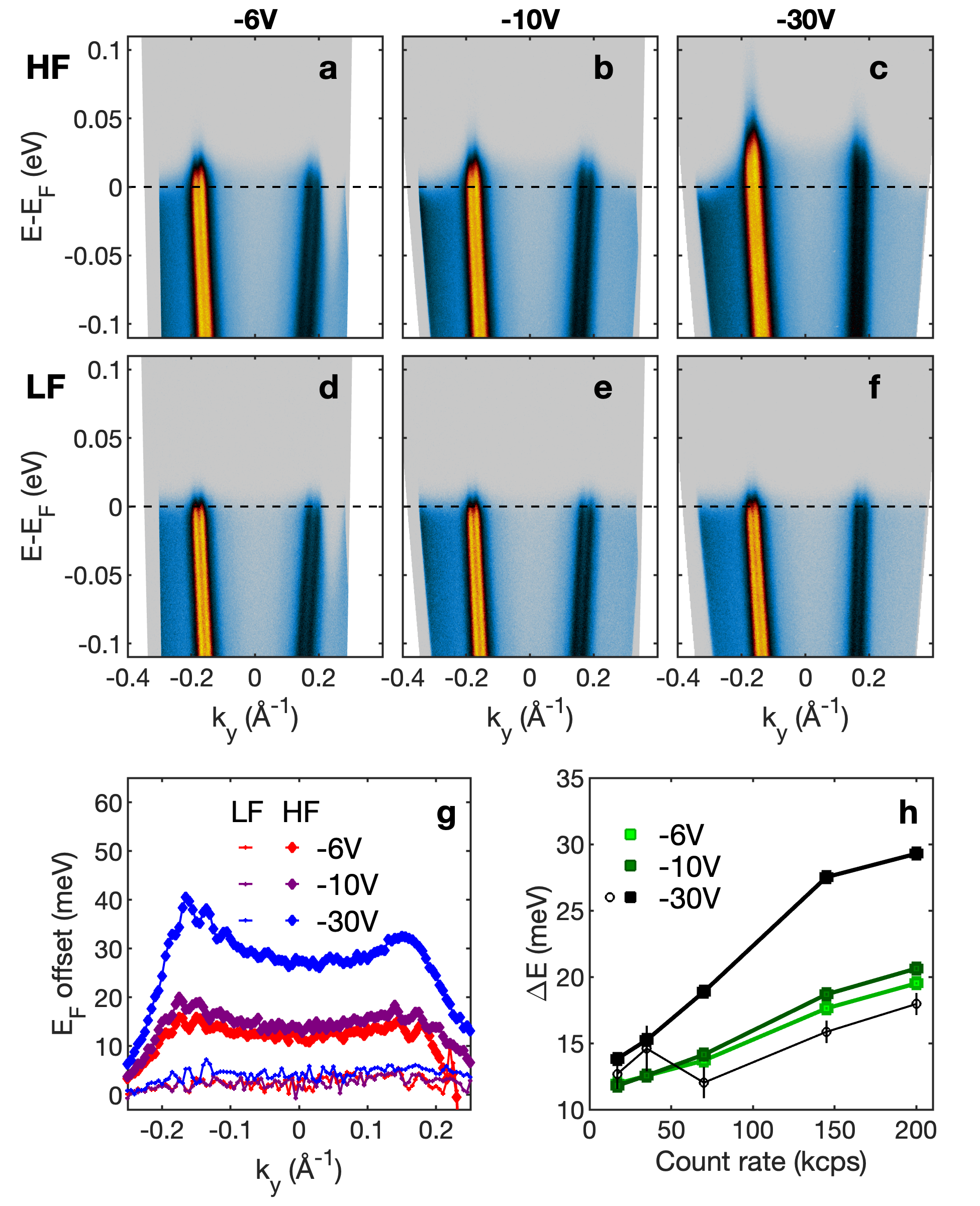}
\caption{(a-c) Au(111) spectra along $k_y$ for $U_B=-6$, $-10$ and $-30$~V in the high flux (HF) regime, showing an enhancement of space charge effects at the largest bias voltage. (d-f) Equivalent Au(111) spectra in the low flux (LF) regime, with negligible voltage dependence. (g) Fermi level $E_F$ shift as a function of bias voltage in the LF and HF regimes. (h) Energy resolution measured at $k=-0.17$~\AA$^{-1}$ (filled squares) as a function of count rate for different voltages. The energy resolution obtained at $k=0$~\AA$^{-1}$ with $U_B=-30$~V is also shown in open circles. The reported count rate is measured at $U_B=-30$~V and the light flux is kept constant for other $U_B$ values.
}
\label{fig:flux}
\end{figure}

Additionally, we report enhanced space charge effects with increasing voltage bias for a fixed flux of the light source.
{Space charge effects are commonly encountered in ARPES experiments when the photoemitted electron cloud is dense, leading to energy shifts and broadening.\cite{Zhou2005,Passlack2006,Graf2010,Hellmann2009} It is well established that these effects depend on the light source parameters such as the pulse duration, the spot size and the photon energy, and we here demonstrate that the voltage bias is an additional contributing factor.}
Spectra acquired at $-6$~V, $-10$~V and $-30$ for high-flux (HF) and low-flux (LF) conditions are presented in Fig.~\ref{fig:flux}a-f. In the HF regime, a clear shift and broadening of the Fermi level is observed with increasing bias voltage, whereas the spectrum remains unchanged across all voltages in the LF regime. 

\begin{figure*}
\centering
\includegraphics{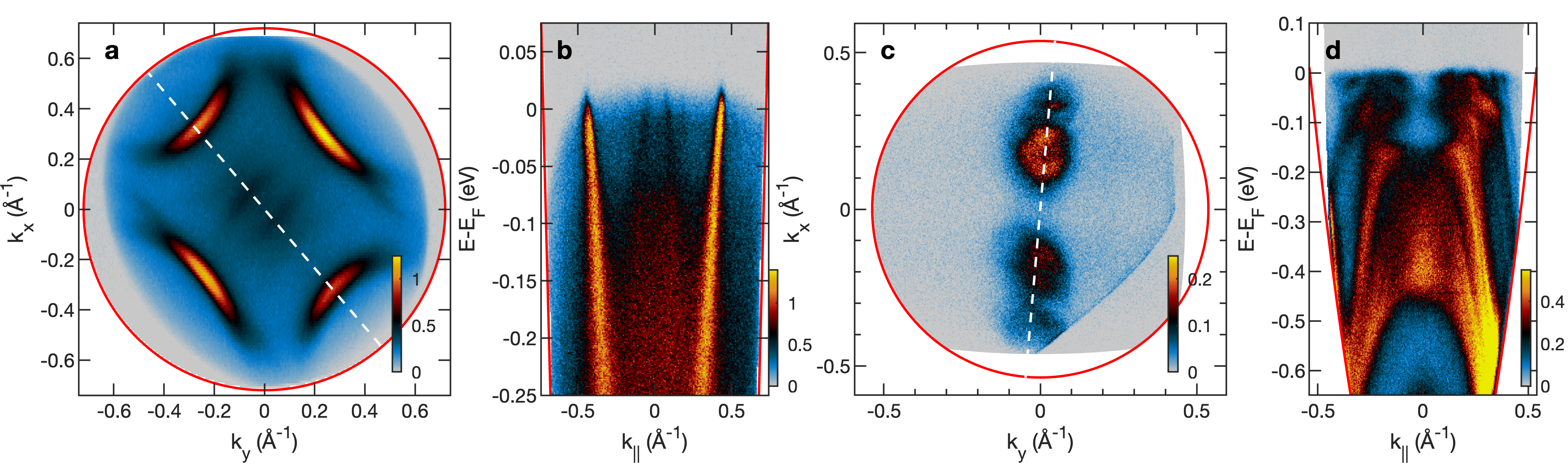}
\caption{(a) Fermi surface of optimally-doped \BSCCO\ measured with $U_B=-48.54$~V. (b) ARPES spectrum along the nodal direction of \BSCCO, exhibiting the weak superstructure replica near $k_\parallel=0$ and the nodes at $k_\parallel\approx0.44$~\AA$^{-1}$. This cut is indicated by the dashed white line in panel a.
(c) Fermi surface of WTe$_2$ measured with $U_B=-20$~V, showing the hole and electron pockets along $k_x$. The sample was misaligned by about 5$^\circ$ relative to the deflector direction. (d) ARPES spectrum along the cut indicated by the dashed white line in panel c, nearly along $k_x$. 
In all panels, the solid red line highlights the physical LEC.}
\label{fig:MaterialExamples}
\end{figure*}

Interestingly, space-charge effect is enhanced in {momentum} regions of high photoemission intensity. This is evident by tracking the Fermi level shift as a function of $k_y$, as shown in Fig.~\ref{fig:flux}g. 
When comparing the Fermi-level shifts for the bands at positive and negative momenta, we note that a larger shift is present for the more intense band at negative momentum. Space-charge effects also lead to an energy broadening, thus affecting the energy resolution. We extracted the energy resolution from EDCs of the band at negative momentum and it is presented {as a function of count rate} in Fig.~\ref{fig:flux}h{. The intrinsic energy resolution is estimated to be 12~meV in the limit of zero count rate, and increases with count rate due to space charge effects}. For $U_B=-6$ and $-10$~V, the dependence on count rate is comparable, while it is more abrupt at $-30$~V, where the enhancement becomes apparent. Similar to the Fermi-level shift, the energy broadening {also} correlates with the photoemission intensity. For example, at $k_y=0$ where the intensity is weak, the space-charge-induced energy broadening is still not visible at $U_B=-30$~V (open circles in Fig.~\ref{fig:flux}h). {This photoemission intensity dependence of space charge is also visible in Fig.~\ref{fig:EDC}c.} Note that all data shown in Fig.~\ref{fig:flux} are along the slit direction, but we also characterized space-charge effects along the deflector direction with similar results.

Finally, we observe that this enhanced space charge effect is not appreciably affected by the slit size. Indeed, we report an energy broadening of only 10\% from 0.1 to 0.5~mm slits at $-30$~V in the HF regime while the detected count rate is more than five times larger. Based on this observation, the space-charge enhancement must originate somewhere between the sample and the slit. 
{To further clarify the origin of the effect, we performed space charge simulations for electrons propagating in a uniform electric field from the sample to the analyzer entrance. The simulation results, presented in Appendix~\ref{appendix:spaceCharge}, show that the Fermi-level offset is enhanced during the electron acceleration by the bias voltage. However, there is no bias-induced enhancement of the energy broadening up to the analyzer entrance. This contrasts with the experimental results, showing enhancement of both the Fermi-level offset and the energy broadening. We therefore speculate that the focus point in the electron lens column of the ASTRAIOS 190 provides an additional contribution to the enhanced space-charge effects.}

\subsection{Applicability for various materials}
\label{subsec:othermaterials}

We demonstrate the capabilities of combining sample bias with deflector technology by presenting results on two well-studied compounds: optimally-doped \BSCCO\ high-temperature superconductor and transition-metal dichalcogenide WTe$_2$. In both cases, we used the weighting factor $\beta=0.85$ established by the characterization of Au(111). The only variable parameter is the sample work function, which is experimentally determined from the LEC to be 4.0~eV for \BSCCO\ and 4.9~eV for WTe$_2$. Based on these work functions, the maximal accessible momentum range at $E_F$ using 6~eV photons is 0.72 and 0.54~\AA$^{-1}$ for \BSCCO\ and WTe$_2$, respectively. In the case of \BSCCO, this corresponds to about 90\% of the tetragonal Brillouin zone boundary. For WTe$_2$, about 60\% of the Brillouin zone boundary is reached along the chain direction, while it is effectively reached along the direction transverse to the chain (although there are no bands at $E_F$ in this momentum space region).

Figure~\ref{fig:MaterialExamples}a shows the Fermi surface of \BSCCO\ obtained with $U_B=-48.54$~V. The four Fermi arcs of the first Brillouin zone are clearly observed, as well as multiple superstructure replicas. As mentioned previously, imperfect alignment can lead to distortions, which appear here as an elliptical shape of the photoemission cone along the $k_x$-$k_y$ diagonal. Here, we present the data using the angle-to-momentum conversion model without further correction, but the ellipticity can be removed by stretching momentum space to either match the LEC circle or fit a known tight-binding model. In addition, an example of an ARPES spectrum taken along the nodal direction is shown in Fig.~\ref{fig:MaterialExamples}b.

Figure~\ref{fig:MaterialExamples}c shows the Fermi surface of WTe$_2$ obtained with $U_B=-20$~V. The electron and hole pockets along the chain direction (nearly $k_x$) are observed. The cutoff in photoemission intensity in the lower-right quadrant is evidence of imperfect alignment. Figure~\ref{fig:MaterialExamples}d presents the rich ARPES spectrum taken along the chain direction. It is noteworthy that this spectrum was measured along nearly $k_x$, which is the deflector direction. The quality of this ARPES spectrum further demonstrates that bias voltage and deflector technology can be effectively combined to push the boundaries of laser-based photoemission. 

\section{Discussion and conclusions}
\label{sec:discussion}

In this work, we demonstrated that combining the bias-voltage approach~\cite{Gauthier2021} with the deflector technology of state-of-the-art hemispherical electron analyzers is a powerful approach for detecting all electrons photoemitted over 2$\pi$ in a fixed geometry in laser-based ARPES experiments. This is especially relevant when using low-photon-energy sources, as their momentum space coverage is intrinsically limited. We extended the previously established model~\cite{Gauthier2021} to consider the deflector axis. We also introduced the weighting factor $\beta$ to account for different lens system designs of hemispherical analyzers. We demonstrated the validity of this extended model for the ASTRAIOS 190 hemispherical analyzer and that a fixed value of $\beta=0.85$ applies to all voltages and samples measured.

In addition to all the advantages and limitations raised previously,\cite{Gauthier2021} the current work showed that the analyzer slit size should be carefully selected to maintain a good transverse momentum resolution. While the slit size affects the transverse momentum resolution even without bias, this effect is notably exacerbated by the application of a voltage that forces photoelectrons with a larger range of transverse momenta through the slit. 
Another important limitation observed in this work is the enhancement of space-charge effects with applied voltage. 
{Space charge simulations demonstrate that part of the enhancement occurs between the sample and the analyzer entrance, but we also speculated that a contribution of this effect arises}
 from the electron lens design{. Further} studies with different hemispherical analyzer designs would be needed to clarify the underlying origin and identify solutions to mitigate this issue. {We note that the extremely large extraction voltages used in momentum microscope also enhance space charge effects,~\cite{Schonhense2018} but novel lens modes are being designed to mitigate this by repelling the low energy electrons.\cite{Schonhense2021,Tkach2025,Tkach2026}}

{As stated in Ref.~\onlinecite{Gauthier2021}, the main experimental limitation of the bias-voltage approach remains the difficulty in realizing an ideal parallel plate geometry with an uniform electric field. One consideration is the experimental design that should be optimized to obtain a symmetric field from the sample surface to the analyzer entrance. In our setup, the flag-style sample holder is inserted in an azimuthal stage, which is electrically decoupled from the manipulator and biased. This biased assembly guarantees a large equipotential surface around the sample. Another consideration is the sample characteristics on field uniformity. Inhomogeneous fields near the surface are induced by rough surfaces, as well as differences in work function of the sample
and its close environment.~\cite{Fero2014} A large and flat sample is preferred to minimize these contributions. 
}

When performing ARPES experiments with low photon energy sources, such as the 6 eV source used in this work, the alignment can be tricky as the photoelectrons have very low kinetic energy and are easily affected by {such field inhomogeneities.}
Advantageously, applying a sufficiently large bias voltage can help reduce the effect of these stray fields. However, as analyzers are generally not designed to be operated with voltage applied to the sample, this can lead to suboptimal electron trajectories in the analyzer, resulting in shifts and broadening in angle and energy.\cite{Dogan2013} While we observed artifacts and distortions, our results do not evidence any significant degradation of instrumental resolution due to the applied bias voltage, as also reported previously.\cite{Gauthier2021}  The principal distortions observed in this work are elliptical Fermi surfaces (Fig.~\ref{fig:MaterialExamples}a), truncated regions (Fig.~\ref{fig:MaterialExamples}b) and kinks in momentum space (Fig.~\ref{fig:slit}a,b). In addition, we observed using the ASTRAIOS 190 analyzer that an imperfect alignment relative to the analyzer's focus distance can lead to a Fermi-level position that depends on the deflector angle. All those measurement artifacts can generally be mitigated by performing a very careful alignment and by using large uniform samples to minimize stray fields.

Considering all those potential distortions, 
one should be extremely careful when combining sample bias and deflector technology to achieve accurate band mapping at absolute momentum values.
The approach is more appropriate in the context of relative studies with a tuning parameter (temperature, pump-probe delay, etc.), or investigations of matrix elements. It is particularly advantageous for TR-ARPES, as various momentum regions can be studied in a fixed geometry, ensuring a constant absorbed excitation density and enabling the comparison of quantitative time-responses across the Fermi surface.~\cite{Armanno2025} 

Our experimental strategy—combining voltage bias with deflector technology—together with its general model, which involves only a single adjustable parameter ($\beta$), should be broadly applicable to a wide range of ARPES systems equipped with state-of-the-art hemispherical analyzers. 
Its applicability is corroborated by an implementation of the same experimental approach recently reported in the literature.\cite{Miao2025} In addition, our work identifies potential limitations (slit-related resolution effects, enhanced space charge, and alignment challenges), thereby facilitating its adoption by other ARPES groups.

\begin{acknowledgments}
We thank the ALLS technical team for their support in the laboratory. The work at ALLS was supported by the Canada Foundation for Innovation (CFI) -- Major Science Initiatives. We acknowledge support from the Alfred P. Sloan Foundation (F.B), the Natural Sciences and Engineering Research Council of Canada (F.B., F.L.), the Canada Research Chairs Program (F.B.), the CFI (F.B., F.L.); the Fonds de recherche du Qu\'{e}bec --- Nature et Technologies (F.B., F.L.), the Minist\`{e}re de l'\'{E}conomie, de l'Innovation et de l'\'{E}nergie --- Qu\'{e}bec (F.B., F.L.), PRIMA Qu\'{e}bec (F.B., F.L.), and the Gordon and Betty Moore Foundation’s EPiQS Initiative, grant GBMF12761 (F.B., F.L.). The work at BNL was supported by the US Department of Energy, office of Basic Energy Sciences, contract no. DOE-SC0012704.
\end{acknowledgments}
\section*{Author declarations}
\subsection*{Conflict of Interest}
The authors have no conflicts to disclose.

\subsection*{Author contributions}

\textbf{N. Gauthier}: Conceptualization (equal); Investigation (equal); Methodology (lead); Formal Analysis (lead); Software (lead); Visualization (lead); Supervision (equal); Writing/Original Draft Preparation (lead); Writing/Review \& Editing (equal);
\textbf{B. K. Frimpong}: Investigation (equal); Methodology (supporting); Formal Analysis (supporting); Software (supporting); Writing/Review \& Editing (equal).
\textbf{D. Armanno}: Investigation (supporting); Formal Analysis (supporting); Software (supporting); Writing/Review \& Editing (equal).
\textbf{A. Jabed}: Investigation (supporting); Writing/Review \& Editing (equal).
\textbf{F. Goto}: Investigation (supporting); Writing/Review \& Editing (equal).
\textbf{V. Hasse}: Resources (equal); Writing/Review \& Editing (equal).
\textbf{C. Felser}: Resources (equal); Writing/Review \& Editing (equal).
\textbf{G. Gu}: Resources (equal); Writing/Review \& Editing (equal).
\textbf{H. Ibrahim}: Funding Acquisition (equal); Resources (equal); Writing/Review \& Editing (equal).
\textbf{F. Légaré}: Funding Acquisition (equal); Resources (equal); Writing/Review \& Editing (equal).
\textbf{F. Boschini}: Conceptualization (equal); Formal Analysis (supporting); Funding Acquisition (equal); Resources (equal); Project Administration (lead); Supervision (equal); Writing/Original Draft Preparation (supporting); Writing/Review \& Editing (equal).

\section*{Data Availability Statement}
The data that support the findings of this study are openly available from the Federated Research Data Repository.\cite{FRDR}
\section*{References}
%


\appendix

\section{{Momentum conversion in the angular and position limits}}
\label{appendix:kconversion}
{This section summarizes the equations and reasoning from Ref.~\onlinecite{Gauthier2021} to obtain the angle-to-momentum conversion in the angular and position limits.}
\subsection{{Angular limit}}
{In the angular limit, we assume that the detected angle is the same one as at the analyzer entrance
\begin{equation}
\Theta_D = f (\Theta_A) = \Theta_A.
\end{equation}
For a parallel plate capacitor geometry, the in-plane momentum is conserved from the sample surface to the analyzer entrance ($k_\parallel^S=k_\parallel^A$) and it is straightforward to write $k_\parallel^S$ in terms of the detected energy and angle
\begin{equation}
k_\parallel^S=\frac{1}{\hbar} \sqrt{2mE_K^D} \sin \Theta_D
\end{equation}
since $\Theta_A=\Theta_D$ and $E_K^A=E_K^D$.
We can rewrite this equation in terms of the kinetic energy at the sample surface
\begin{equation}
k_\parallel^S=\frac{1}{\hbar} \sqrt{2mE_K^S} F_A \sin \Theta_D
\end{equation}
with $F_A = \sqrt{1+2\alpha}$ and $\alpha$ is defined in Eq.~\ref{eqAlpha}.
}
\subsection{{Position limit}}
{
In the position limit, we assume that the detected angle $\Theta_D$ is only a function of the arrival position on the analyzer entrance $x_A$ (see Fig.~\ref{fig:Schematic}d). This implies 
\begin{equation}
\Theta_D=f(x_A)=\arctan{(x_A/d)}
\end{equation}
where $d$ is the sample to analyzer distance. Considering an uniform electric field between the sample and the analyzer entrance, the position $x_A$ is obtained from basic kinematics, 
\begin{equation}
x_A=\frac{v_x}{a} \left( \sqrt{v_z^2+2ad} -v_z \right), 
\end{equation}
where $\vec{v}=(v_x,0,v_z)$ is the initial velocity and $a=-eU^*_B/md$ is the acceleration along the $z$-axis. Here, we assumed $\Omega=0$ for simplicity but this is generalizable for any azimuthal angle. Combining both equations and rewriting $v_z$ in terms of $E_K^S$, we obtain
\begin{equation}
\tan \Theta_D = \frac{v_x}{ad} \left( 
\sqrt{\frac{2E_K^S}{m}-v_x^2+2ad}
-
\sqrt{\frac{2E_K^S}{m}-v_x^2}
\right).
\end{equation}
This equation is solved for $v_x$ and only the solution that remains finite in the limit $a \rightarrow 0$ is kept. Rewriting in terms of $k_\parallel^S=mv_x/\hbar$, we find
\begin{equation}
k_\parallel^S=\frac{1}{\hbar} \sqrt{2mE_K^S} F_P \sin \Theta_D
\end{equation}
with
\begin{equation}
F_P= \sqrt{\frac{\alpha+1+\sqrt{2 \alpha + 1 - \alpha^2 \tan^2 \Theta_D}}{2}}.
\end{equation}
}

\section{Scaling of photoemission intensities}
\label{appendix:IntensityScaling}

The scaling of photoemission intensities due to the change of coordinates is given by 
\begin{equation}
\frac{N(E_K^S,k_x^S,k_y^S)}{\Delta k_x^S \Delta k_y^S}=
\frac{N(E_K^S,\theta_X,\theta_Y)}{\Delta \theta_X \Delta \theta_Y} \left| J_{\Theta_D,\Omega_D}^{\theta_X,\theta_Y} \right|^{-1}
\end{equation}
where the Jacobian is
\begin{equation}
    J_{\Theta_D,\Omega_D}^{\theta_X,\theta_Y}= 
    \begin{bmatrix}
    \frac{\partial k_x^S}{\partial \theta_X} & \frac{\partial k_x^S}{\partial \theta_Y} \\
    \frac{\partial k_y^S}{\partial \theta_X} & \frac{\partial k_y^S}{\partial \theta_Y}
    \end{bmatrix}.
\end{equation}
All relevant partial derivatives to write the Jacobian are expressed below. 

\begin{multline}
\frac{\partial k_x^S}{\partial \theta_{X,Y}}= \frac{1}{\hbar} \sqrt{2mE_K^S} 
\bigg( 
\frac{\partial F}{\partial \theta_{X,Y}} \sin \Theta_D \cos \Omega_D  \\
+F \cos \Theta_D \frac{\partial \Theta_D}{\partial \theta_{X,Y}} \cos \Omega_D \\
-F \sin \Theta_D \sin \Omega_D  \frac{\partial \Omega_D}{\partial \theta_{X,Y}} 
\bigg)
\end{multline}
\begin{multline}
\frac{\partial k_y^S}{\partial \theta_{X,Y}}= \frac{1}{\hbar} \sqrt{2mE_K^S} 
\bigg( 
\frac{\partial F}{\partial \theta_{X,Y}} \sin \Theta_D \sin \Omega_D  \\
+F \cos \Theta_D \frac{\partial \Theta_D}{\partial \theta_{X,Y}} \sin \Omega_D \\
+F \sin \Theta_D \cos \Omega_D  \frac{\partial \Omega_D}{\partial \theta_{X,Y}} 
\bigg)
\end{multline}
%
%
\begin{equation}
\frac{\partial \Theta_D}{\partial \theta_{X,Y}}= \frac{\left(\tan^2 \theta_X + \tan^2 \theta_Y\right)^{-1/2}}{1+\tan^2 \theta_X + \tan^2 \theta_Y}  \frac{\tan \theta_{X,Y}}{ \cos^2 \theta_{X,Y}}
\end{equation}
\begin{equation}
\frac{\partial \Omega_D}{\partial \theta_X}= \frac{1}{1+{\frac{\tan^2 \theta_X}{\tan^2 \theta_Y}}} \frac{\sec^2 \theta_X}{\tan \theta_Y}
\end{equation}
%
%
\begin{equation}
\frac{\partial \Omega_D}{\partial \theta_Y}= \frac{-1}{1+{\frac{\tan^2 \theta_X}{\tan^2 \theta_Y}}} \frac{\tan \theta_X}{\sin^2 \theta_Y} 
\end{equation}
\begin{equation}
\frac{\partial F}{\partial \theta_{X,Y}}= \frac{-\beta \alpha^2 \tan \Theta_D \sec^2 \Theta_D}{4 F_P \sqrt{2 \alpha + 1 - \alpha^2 \tan^2 \Theta_D}} \frac{\partial \Theta_D}{\partial \theta_{X,Y}}
\end{equation}

\section{{Simulations of bias-enhanced space charge effects}}
\label{appendix:spaceCharge}
{
Space charge simulations were performed to clarify the origin of its enhancement under applied bias. The simulations are inspired by the ones reported in Ref.~\onlinecite{Zhou2005}. A set of $N_e$ interaction electrons is generated assuming a uniform density of states with no momentum dependence, emitted with a kinetic energy between 0 and 0.5 eV, corresponding to the photoemission from a 5.5 eV workfunction material [such as Au(111)] with a 6~eV light source. The $N_e$ electrons are sampled over the spot size and pulse duration by considering Gaussian spatial and temporal distributions with FWHM of 100~$\mu$m and 300~fs, respectively. The electrons are propagated from the sample surface to the analyzer entrance, over a distance of 30~mm in the presence of a uniform electric field. The interaction electron propagation is described by analytical kinematics and all electron-electron interactions between those electrons are neglected. The space charge effects are simulated by evaluating how one test electron is affected by all the $N_e$ interaction electrons.
}

\begin{figure}
\centering
\includegraphics{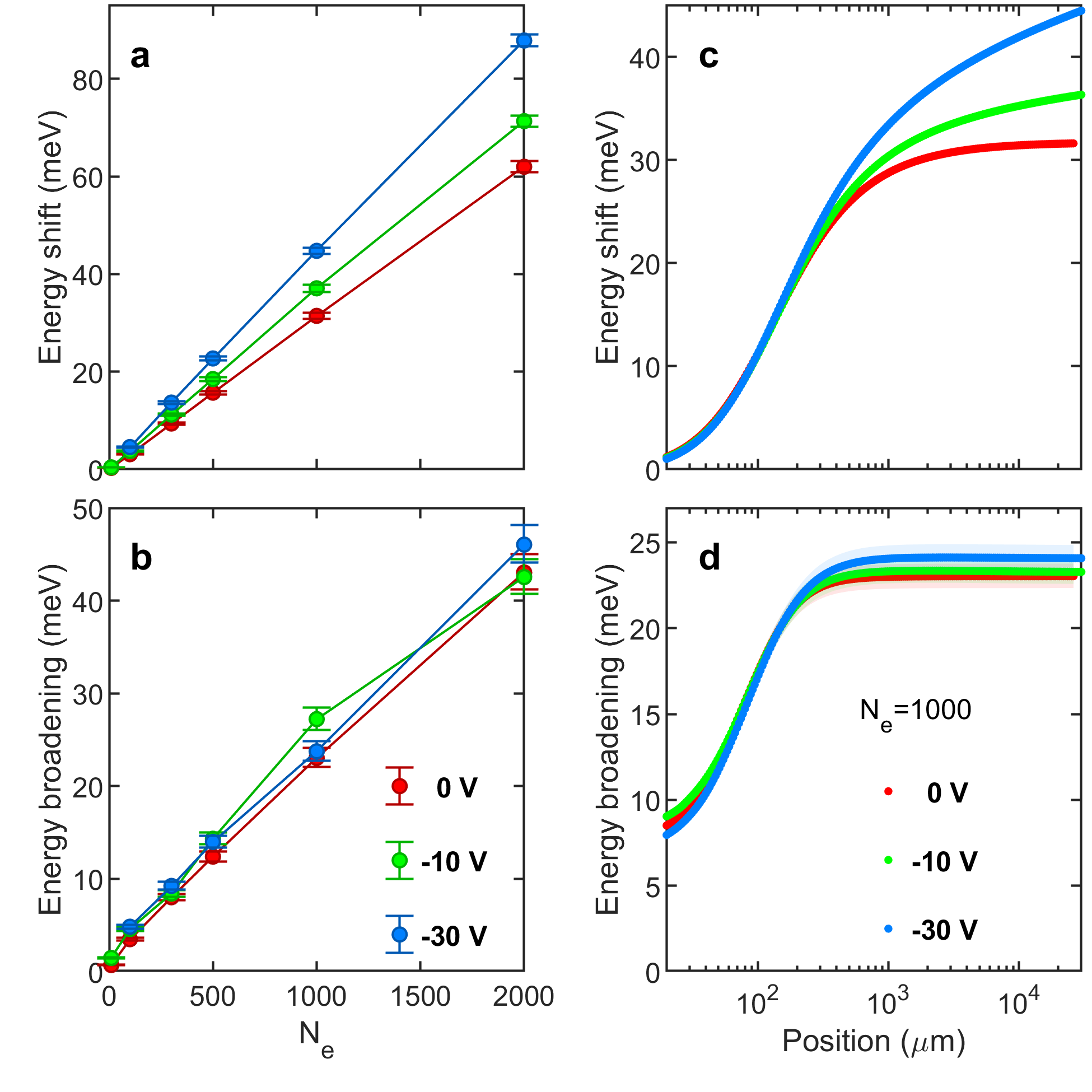}
\caption{{Space charge simulations as a function of bias voltage. (a) Fermi-level offset and (b) energy broadening as a function of the number of interaction electrons $N_e$ for different bias voltages. (c) Energy shift and (d) energy broadening accumulated over the test electron propagation from the sample surface (0~$\mu$m) to the analyzer entrance ($3 \times 10^4$~$\mu$m) for $N_e=1000$. The position is taken along the sample-to-analyzer direction ($z$-axis).}} 
\label{fig:spacechargesim}
\end{figure}

{
The test electron is photoemitted with $E_K=0.5$~eV (at $E_F$) at normal emission ($\Theta=0^\circ$) within the spot size and pulse duration. Its propagation from the sample to analyzer entrance, considering the force exerted by all $N_e$ electrons and the uniform electric field, is evaluated numerically. This numerical procedure is repeated $n_\text{repetition}=1000$ times and the Fermi level offset and broadening is estimated from the mean and FWHM of the obtained distribution. 
As discussed in Ref.~\onlinecite{Zhou2005}, mirror charges of the interaction electrons are expected in front of a metallic surface and can affect the space charge effects. We performed simulations with and without mirror charges and the results are qualitatively the same, with only changes in magnitudes. Fig.~\ref{fig:spacechargesim} presents results with mirror charges included.  
}

{
The space charge effects on the energy shift of the Fermi level and the energy broadening are presented in Fig.~\ref{fig:spacechargesim}a-b as function of the number of interaction electrons $N_e$, for various bias voltages. For a fixed $N_e$, the energy shift increases for larger $|U_b|$ while the energy broadening is unchanged. This dichotomy can be understood by tracing the accumulated energy shift and broadening along the test electron propagation, as shown in Fig.~\ref{fig:spacechargesim}c-d. As already noted in Ref.~\onlinecite{Zhou2005}, the energy shift increases gradually while the broadening happens much faster. In simple terms, the broadening occurs from random fluctuations when the electron cloud is dense while the energy shift is resulting from the constant electron cloud expansion. In our simulations, the broadening increases only within about 200~$\mu$m of the sample surface and remains constant afterwards, for all bias voltage values (Fig.~\ref{fig:spacechargesim}d). Within such a short distance, the electron acceleration from the bias voltage is negligible with a gain in kinetic energy of less than 1\% of $|U_b|$. This justifies why broadening effects are not enhanced by bias voltage of modest values (tens of V). In contrast, the energy shift occurs over a larger distance (Fig.~\ref{fig:spacechargesim}c) and acquires a voltage-bias dependence.}

\end{document}